# Augmenting a Firefighter's PPE - Gas Mask (SCBA)

Kunal Aneja
Undergraduate Student at Georgia Tech
+1 (770) 570-7670
kaneja6@gatech.edu

Tejaswini Ramkumar Babu
Undergraduate Student at Georgia Tech
972-695-8200
tbabu8@gatech.edu

Rachel Chan
Undergraduate Student at Georgia Tech
+1 (615) 481-8714
rchan40@gatech.edu

## ABSTRACT
PPE (Personal Protective Equipment) has allowed firefighters to perform their everyday tasks without getting harmed since the mid 1800s. Now, the advancement of technology has given rise to the improvements of PPE. PPE can now include sensors to detect any number of environmental hazards (chemical, biological, temperature etc.). As the GT class of CS3750, we have decided to create a version of an interface design sensor that will help firefighters in two ways: navigation and communication. In order to augment a firefighter display when they are within a building, we chose to augment their SCBA (self-contained breathing apparatus). The gas mask will include a small screen that displays vital information directly towards the firefighter without need of any other support. We used the Google Glass to display vital information directly towards the eye in a minimalistic manner, while also augmenting that by adding LED lights to simulate someone calling their name or other auditory signals. While our prototype focuses on two main components of a firefighters search and rescue in a building, both of them combine to augment a firefighters display when searching throughout a building to help improve accuracy, speed and overall experience.

## CCS Concepts
• Human-centered computing ~ Visualization ~ Visualization application domains ~ Information visualization

## 1. INTRODUCTION
We will be researching the implications of having a heads-up display on the helmets of first responders for navigational use. This will be a quick, hands-free way to notify them of the location of other first responders and other key information based on the situation. As the GT class of CS3750, we have decided to augment current firefighter PPE (Personal Protective Equipment) by creating a version of an interface design sensor that will help firefighters in two ways: navigation and communication. Firefighters need to wear PPE to protect themselves from various dangers while on duty while doing their tasks. Now, the advancement of technology has given rise to the improvements of PPE. PPE can now include sensors to detect any number of environmental hazards (chemical, biological, temperature etc.). Thus, we have designed a heads up display that will augment a firefighter's field of vision, located in the periphery of their SCBA (self-contained breathing apparatus). The gas mask will include a small screen that displays vital information directly towards the firefighter without need of any other support. The design will allow the user to see where other teammates are located, communicate spatial information, and direct the firefighter's attention to specific audio signals by using LED lights. The design displays vital information directly towards the eye in a minimalistic manner. While our prototype focuses on two main components of a firefighters search and rescue in a building – navigation and communication – both of them combine to augment a firefighters display when searching throughout a building to help improve accuracy, speed, and safety.

## 2. TASK ANALYSIS
There are a variety of different assignments firefighters have, of which include survivor extraction during a major fire situation. This process includes the preparation of equipment, wearing PPE, transportation to the location, determining the game plan, entering the building in the methodical manner, running a quick priority one search based on known layouts of building in the area, and then a later priority two search with a more detailed search for any known survivors. If a firefighter found a survivor, the survivor becomes a patient and the firefighter must run an EMT analysis on the patient which includes checking for injuries and life-threatening conditions other than the fire. Furthermore when a firefighter finds a survivor an analysis on survivability must be done which includes the chance to extract the survivor alive without injuring themselves and "the challenge is that empathy and compassion are inherent to caregivers."

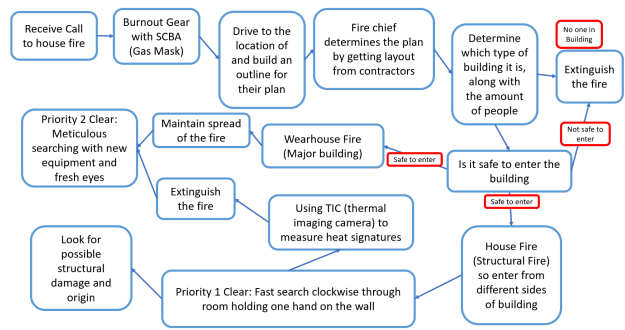

*Fig 1. The method a firefighter goes through when receiving a call for search and rescue within a fire engulfed building.*

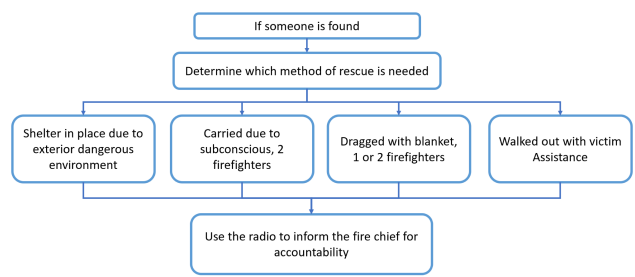

*Fig 2. The method a firefighter goes through after finding a survivor within a burning building*

# 3. EVALUATION TECHNIQUES AND TASKS

## 3.1 Navigation Section

To augment a firefighter's navigation aspect, we aimed to create a HUD (Head-Up Display) within a firefighter's SCBA (Self-Contained Breathing Apparatus) which displays the movement and position of his/her other crew members. When creating this prototype, we had to assess how we would demonstrate the prototype. We created a video demonstration using Imovie, Ibis Paint, and Final Cut Pro so that the participant can watch to view how our interface would be implemented into their gas mask. Before we displayed the demonstration, we evaluated the navigation component through questioning for qualitative information. Later, the participant first watched a video demonstration of a firefighter's body camera with a navigation map on the top left corner. There's a navigation animation on the top left corner that will route out the paths of the firefighters from the video by showing each of the firefighter's locations as a dot on a 2D map.

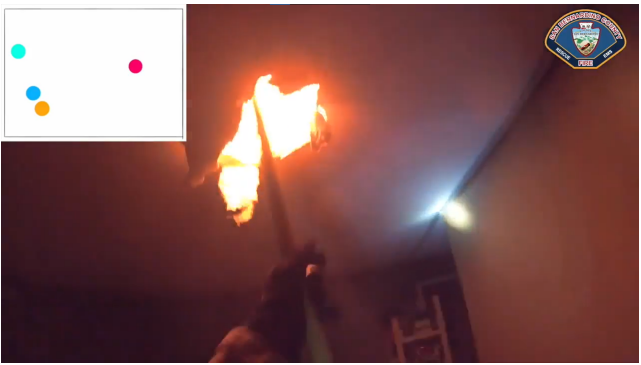

*Fig 3. An example frame of the video displayed to firefighters when running the navigation trial.*

While the participants watched the video, they relayed to us comments we asked of them. For instance, we asked "do you see any movement on the map, can you tell us where "this color" is going?" After watching this demonstration, we presented the tester with google glasses. The participant wore the google glass and watched our simulation through the glasses, our low-fidelity prototype for our HUD in the SCBA gas mask. These google glasses act closely to how the interface will be displayed on the real gas mask. When the participant put it on, they were asked about the comfortability of the google glass and whether the interface is too big, where it would be destructive of the interface, and if the map was too large or too small. Once the participant had familiarity and feel over the google glass, the navigation map animation was displayed on the google glass. We then asked the participant how distracting the interface was if the navigation map was too big or too small, and the location where they would prefer the navigation map. Furthermore, we took our notes based on the situation. Next, we showed the participants the other two map sizes to see which ones they preferred. We also showed the participants the second navigation video design in three different sizes, which had a black background in which the dots had trails instead of a white background. Lastly, we gave all participants a NASA TLX test to measure the amount of cognitive and physical effort it took the participants to do these tasks.

## 3.2 Communication Component

In order to augment a firefighter's communication aspect we used color signals detailing who is trying to signal him and the information. We tested using the LED lights to help notify the firefighter in high cognitive environments, and recorded and compared the tests with different types of color and flashing. For the demonstration, the participant will be shown a video of a firefighter cam that is going through their operation. As the participant watches the video he will see at some points in the video, there will be a subtle red or blue color flash that will signify who is contacting them. When the fire chief calls out on the radio, the firefighters name and the screen turns blue to signify a callout. When one of his crewmates is trying to signify their name or major yelling is heard, the screen will turn red to signify the callout.

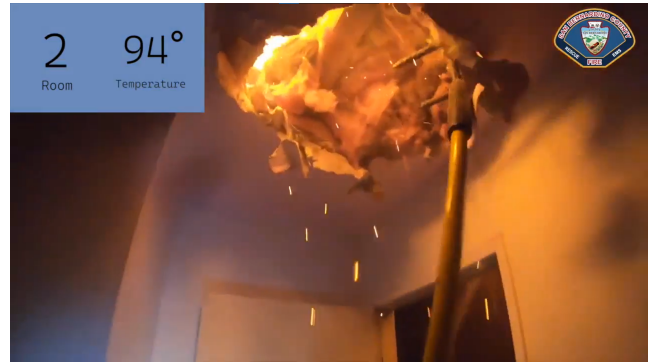

*Fig 4. An example frame of the video displayed to firefighters when running the communication trial.*

The demonstration video was done through Figma in order to create a layered interface design video. In addition to Figma, Anima was used as a plugin to incorporate the video into the display. Through this demonstration we will ask him to call out what it means when the color flashes blue or red and when he sees his name, if the flash is too fast, if the colors are too bright, etc. As for the actual prototype, we created an attachment that could be added to the google glass.

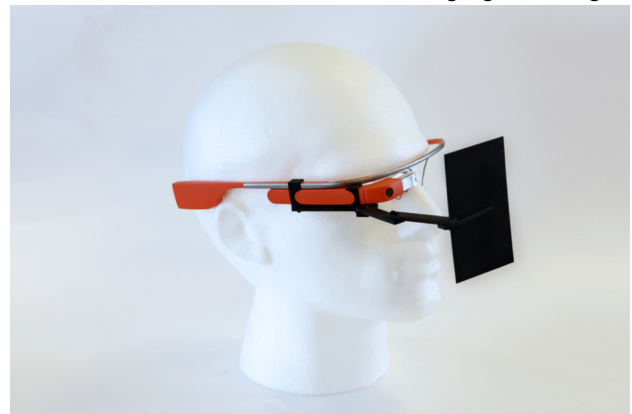

*Fig 5. A 3D model of the printed extension device to the Google Glass for the LED additions.*

The attachment was 3D printed using the resources at Georgia Institute of Technology (GT), and LED lights were soldered together, and were wired up through the back of the glass. The attachment is controlled with a switch to signal the red and blue flash. As the participant is wearing the glasses with the attachment, the participant will be asked about the comfortability of the glasses during wearing, whether it affects vision while wearing a mask, the amount of focus he had on the task at hand, and whether it will improve and augment his use case. After watching the videos and hearing the audio feedback, the participants were asked questions about distraction level, and then given the NASA TLX survey again.

## 4. RATIONALE

We used the videos and the google glass as a low fidelity prototype of a heads up display in a SCBA. This way, we were able to emulate a peripheral interface. When watching the video, the firefighters will be able to put themselves in the first-person point of view of the firefighter in the video. The participant will pretend that he/she is the firefighter in that situation, just as he/she would in real life when using the prototype. Though the participant won't have control of the actions done by the person in the video, the participant will be able to see his/her location based on the navigational map on the left side of the screen in the user's peripheral vision.

During the video, the participant was asked about the actions in the video, to see if he/she can recognize the bodycam, see what the firefighter in the video is doing, his/her field of view, and what the dots are doing. These questions were asked during the video in real-time so we could better understand how easy it was for the participant to grasp the technology and understand what was going on, especially since a large aspect of our HUD technology is accessibility. The questions that ask about what tasks the firefighters in the video are doing is to make sure the participant understands what is being done, so the participant can imagine himself/herself as the person doing the tasks in the video. Furthermore, these questions allow us to determine what aspects of the interface that the participant is focusing on during each of the trials. This way, we can find which aspects are the most important to a firefighter during an emergency situation.

After the video, the participant was asked where the crew members come together in terms of the positioning from the animation, what he/she most focused on in the video, which of the three sizes he/she preferred the most, and the number of times he/she looked at the top left where the map was. These questions are to determine how well the firefighters understand and use the interface. Their detailed explanations after their experience with the trials will provide more insight into what the firefighters think about the usefulness about the device.

The above tasks were repeated with the second navigation map video that had a black background instead of a white background. The above trials were then repeated again, with the same three different sizes of maps. The video with the black background was shown to the participant to determine how well the black background worked better or worse than the white background. Furthermore, trails were added to the dots that indicate where each of the firefighters had been. The trails showed the previous location of the firefighters, and the participants were able to describe if the trails were distracting or not. This way, the most functional and easy-to-read map based on color and type of location indicator (dots or trails) can be determined to create the best kind of navigational map.

For the communication aspect, one video was played that displayed data, gave firefighters audio feedback from the fire chief, and lit up different colors based on who was talking. We wanted to see if the data distracted or helped the firefighter. Additionally, the different colors signal if the fire chief or another is talking, and lets the firefighters know that they should prioritize their attention. However, these trails were done to make sure the users are not distracted by the colors. By determining their feedback, we will know if these features are distracting, if they are helpful, and what can be improved on.

The NASA TLX survey allows us to measure the participant's attitude on the amount of cognitive or physical effort it took to do these tasks we presented to him or her. We used the data collected from the NASA TLX to study comfortability and wearability to adjust our prototype to match the most common range of the circumferences of the users' heads.

## 5. PARTICIPANTS

We interviewed and complemented the trails with two firefighters and a police officer. Our heads up display technology is meant to be mounted inside a firefighter's SCBA mask and our demonstration video is of a firefighter trying to keep a fire from spreading, so this experiment is designed for participants who are experienced firefighters. However, the police officer is a supplemental participant who provided valuable secondary feedback on our device and its feasibility, as someone with years of knowledge in being a first responder.

### 5.1 Firefighter J

Firefighter J is a firefighter, but he does more than just fighting fires. He speaks to students at school about enhancing fire safety, checks water hydrants, and provides emergency medical services. During his down time at the station (times that he is on shift but there is no call come in), he does lots of high end training to maintain his skills as it is a life-or-death matter. He understands that his job is stressful at times, but he is glad he can help people.

### 5.2 Firefighter X

Firefighter X has been a firefighter for over seven years, five of which have been in the Atlanta Fire Rescue Station 15. She must be prepared whenever a call comes, 80% of which are medical and 20% are fire. If it is a fire call, she changes into bunker gear, or turnout gear, and gets her team ready. Each firefighter is connected to the radio where the fire chief is directing everyone around the building. When searching through a building, she follows a specific pathway mapping the area in order to most efficiently clear out the building. When she finds a survivor, the details are relayed back to the fire chief through the radio and the fire chief gives further instructions to deal with the situation.

### 5.3 Police Officer Y

Georgia Tech Police Officer Y has been a police officer for over 25 years and has entered a building, which simulates this experience over 200 times. While this device was designed for a firefighter's gas mask, police officers work in similar scenarios when entering a building and are required to clear every room, for example an active shooter scenario. Police officer Y works with a team to clear every room with a task in mind while connected through radio to dispatch, which are similar situations as firefighters.

## 6. RESULTS OF THE STUDY

### 6.1 Navigation

The overall consensus of the heads up display was highly positive and improved their experience, however there were many requirements and changes required before deployment. With each participant ranging in search and rescue experience, many of the changes were software based with each adjusted to requirements of each user. It took a total of 30-40 minutes for each participant.

Some of the situations where the HUD would be exceedingly practical would be when many firefighters are moving a water-hose around a building. By recognizing where each firefighter is located around the building the movement and transportation time of the hose can be greatly decreased. Furthermore, by positing a firefighter at each turn in the hose can prevent common risks and mistakes such as bending or dragging. As for other first responders a spatial understanding can be helpful in an active shooter scenario because one can see where your backup is and inform others of directions.

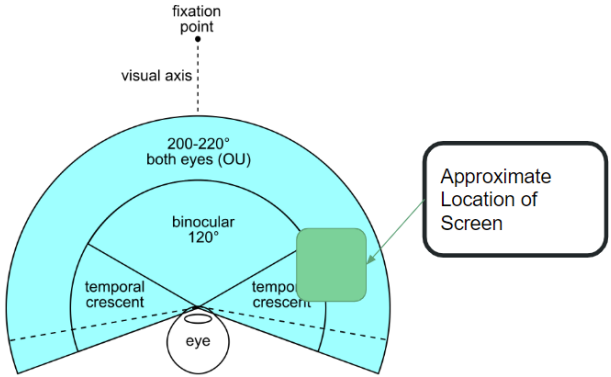

*Fig 6. A demonstration of a firefighter's field of view when wearing a SCBA and the approximate location of the screen in the peripheral vision.*

As the display was located in the corner of the mask it didn't deter the task at hand and wasn't a distraction. According to the figure above, out of the 200 degrees of vision a firefighter's mask covers the temporal crescent leaving only 120 degrees of movement. However when completing a task the firefighter's main focus is on the center 80 degrees of the binocular view. This proved that the device was located in a perfect position. Moreover, 90% of the time the bottom 3/4 of the mask is used, especially when crawling and in low visibility environments.

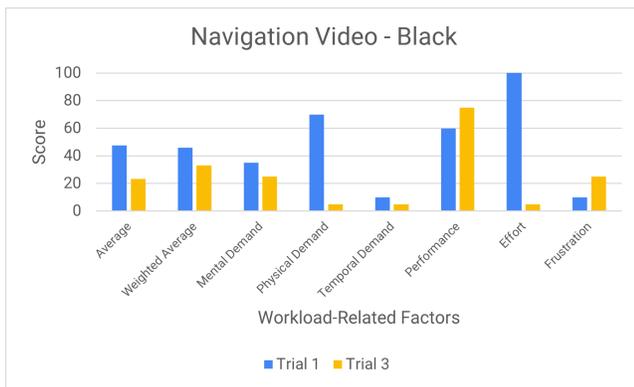

*Fig 7. The NASA TLX analysis of the navigation trial with the black background.*

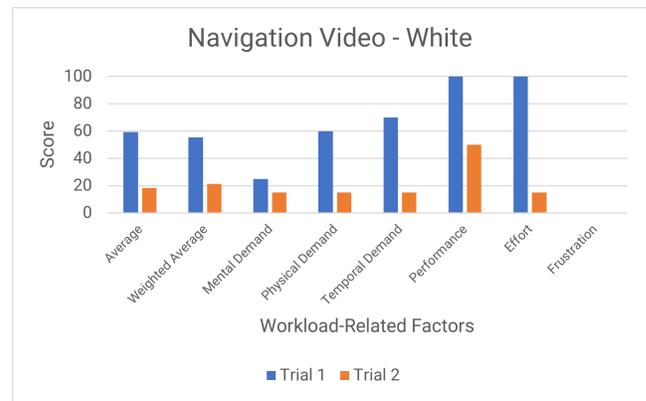

*Fig 8. The NASA TLX analysis of the navigation trial with the white background.*

As shown in the two figures above, the black background for the navigation trial was prefered. This was because the regular environment this device would be used in would be much darker due to the smoke, so having a screen that doesn't overly stimulate the eyes would be preferred. Furthermore, the trails indicating the previous movements of the firefighters were not chosen due to the excess amount of unnecessary information. The weighted average of the NASA TLX survey shown above demonstrates this idea. The weights for each section was predicted based on known firefighter tasks in search and rescue environments.

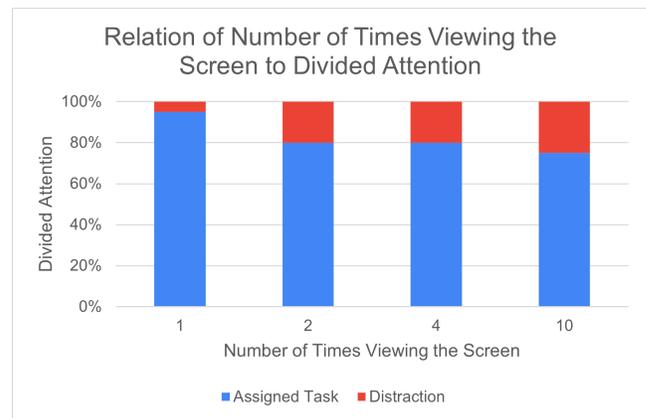

*Fig 9. The trend of lowered rate of focused attention on the task in relation to the number of times viewing the screen.*

As shown in the graph above, Relation of Number of Times Viewing the Screen to Divided Attention, the device has limited distraction over a period of time and was only viewed when necessary. For example when moving between rooms or moving equipment (like a water hose) throughout the building. Due to the limited detailed information shown on the display, the size of the screen required to be minimized as many firefighters don't require extremely detailed location information, just the general idea.

The device helps give awareness and understanding of the map and its relation to the real world, for example being able to comprehend where teammates are through the walls, essentially giving themselves a much sought after feature, x-ray vision.

### 6.2 Communication

For the communication aspect, the consensus was that this feature must be highly minimized to prevent obvious distractions;

however, it can improve communication time by 40%. When a firefighter is within a building, the fire marshal or battalion chief is positioned outside the building receiving information from contractors and other first responders to direct his team around the building. The communication is done over an outdated technology, radio. The radio chirps before coming on and is loud enough for others nearby to hear it.

During the period of an hour within a building each firefighter uses the radio 3/4 times and out of that 40-50% times must be repeated due to white noise and this device can bring it down to a predicted 5%.

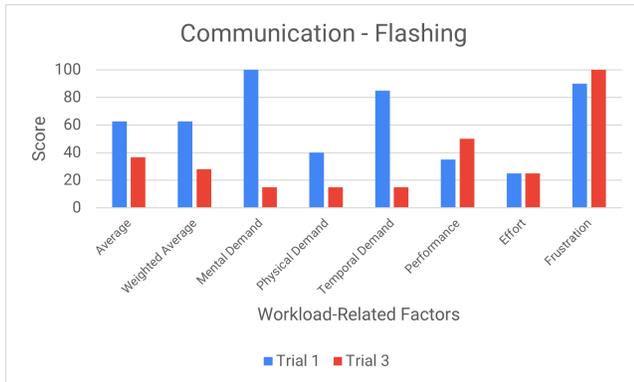

*Fig 10. The NASA TLX analysis of the communication trial with the flashing screen.*

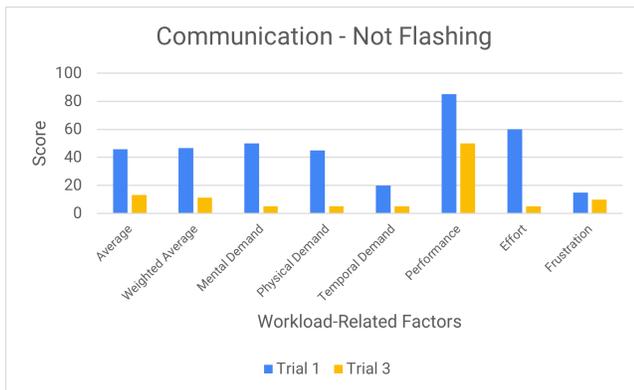

*Fig 11. The NASA TLX analysis of the communication trial without the flashing screen.*

In one of the scenarios flashing was highly discouraged and only when high priority information was being shared. An example is when a priority 1 scenario comes up like a key patient or fire has become out of control. This is shown in the figure above on the high levels of frustration and temporal and mental demand usage. On the other hand the information displayed should always be included because in high energy scenarios people always forget room numbers and this is a quick way to check that out.

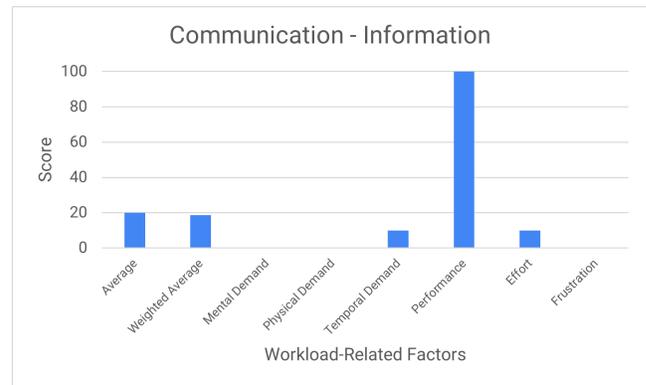

*Fig 12. The NASA TLX analysis of the communication trial with the information displayed.*

Finally shown in the graph above, performance was highly improved when the detailed information was displayed and not only the colors. This was shown in the trials when the firefighter forgot the information and was assisted with the information displayed in the screen allowing them to continue their tasks.

## 7. IMPLICATIONS OF RESULTS

Overall the device was shown to improve speed and accuracy of the many tasks completed by firefighters but due to the new technology further testing and training is required before deployment.

Using the HUD requires previous knowledge and training in order to position the device and inform the type and quality of information displayed. Nonetheless, without training firefighters, they noticed dots and their movement and how it represents themselves and other teammates. Some failed to locate which dot represents themselves so they requested an option to lock the map to rotate as you rotate giving a first-person map interface or to have a triangle represent themselves. In addition some other developments include integrating a 3-D map with a graphical interface to best correlate the building to map rather than the current 2D map.

An unpredicted situation that came up was how firefighters could become dependent on the device causing them to ignore the fire marshal speaking and only pay attention to the display causing them to miss vital information. This can become detrimental if there is a software error which displays the wrong information. This can be fixed by further training on the device and preventive software to prevent a life-threatening event. Another key training aspect is to know which color represents each person talking, which gives a subconscious signal and understanding of the information.

## 8. DESCRIPTION OF TRIAL UI CHANGES

Some of the changes made during the testing process was the adjustment of the display. The google glass had different positioning settings for different use cases so we made real time adjustments to fit the device to the user's head as comfortably as possible. Moreover, we adjusted the size of the screen in real time to best understand the size and position of the screen which would be preferred.

## 9. DESCRIPTION OF CONCLUDED CHANGES

Some of the changes we would make as a team if we had more time is using a real gas mask to implement our prototype, create a HUD (heads up display) within the gas mask instead of the google glass, and interviewing more of the participants targeted to our product.

If our team was able to get our hands on a gas mask, this would make the "try-on" trial of our research more realistic because it closely portrays what PPE we are focusing on for firefighters. In addition to obtaining the gas mask (SCBA), we could also use the extra time to implement a display screen on the top corner of the gas mask. The display screen could be connected by an HDMI cable of the computer to where the computer can play the navigation animation and the communication video from the demonstration video we showed our participants. This would then transfer to the display screen of the gas mask. This would allow the participant's feedback to be more accurate since this interface design would closely resemble how we envisioned it to come out. Even though the google glass was a good fidelity prototype in its use of displaying a screen in the participant's peripheral vision, the wearing and comfortability of the glasses does not strongly resemble how it would be with the gas mask, which in turn, could mess with the quantitative results. Furthermore we hoped to have the google glass adjustable to which side it displays the information. Finally, if we were able to get more of our target participants to give us feedback and participate in our trials, we would be able to receive more results and have an even more robust analysis on our findings of measuring their performance, effort, thoughts, usability, etc.

## 10. LEARNED EXPERIENCE

After wrapping things up with our project, our team has learned a lot of valuable knowledge and lessons concerning the creation part of UI design and the power of UI design. When coming up with an interface design, our team at first would narrow one situation down and come up with a possible solution. However, when creating a product, it is best to list out the different scenarios the target participants go through and whether there might be a lack of information in this area. Our team was narrow minded at first, but with the help of our instructors we were able to come up with different scenarios, problems, and solutions. It is until then can we really decide on what idea to move forward with.

We learned that all three of our members had showcased their own abilities within this project. Kunal was great at identifying the various tasks and procedures firefighters would go through. His knowledge from his research allowed him to give us possible scenarios firefighters would go through when dispatch calls. His researching skills paved the way to our possible designs that we dedicated when helping firefighters out with these scenarios. Teja has the ability to visually and verbally explain ideas well. When either of the teammates are confused, she will give us simple explanations that allow us to understand why this certain design might not be plausible or why firefighters may not want this certain type of signaling when they are going through heavy fire and smoke. Rachel has the ability to record what needs to be picked up. Having a tab of what the team talked about or what the professors talked about is crucial. In order to learn from the feedback or knowledge of what we obtained, we must be able to keep track of our progress so that our prototype will be worth it in the end. Together, we were able to communicate to each other, give roles in one's strengths, and generate conclusions that would help draw us closer to our research questions and method to the development of our prototype. In addition to the development process of UI design, we also obtained more knowledge on UI design. When researching different ways to implement sensors onto the PPE of firefighters, we did not know the types of technology that was already out there. For example, some firefighters useTIC (thermal imaging camera) that can identify areas of different heat levels. In addition, there are many types of sensors that can be implemented to heads up displays. When researching different participant interface design signals, we found there to be a bunch of options: LED color changes, heads up display, and haptic feedback. There were many ways we could implement an interface design onto the PPE of a firefighter. However, we learned that we need to know from firefighters themselves as to what type of feedback sensors would be helpful or not when they are heavily focused on the real task they have been ordered to do.

If we could start over, we would change our way of thinking and become more open minded to different focuses of firefighters. Instead of thinking "navigation" as our only focus, we should have definitely thought of more focuses because like how our prototype ended, we implemented two focuses rather than one. In addition, we would also put ourselves more in the shoes of our participants. From reading The Design of Everyday Things[1], it is great if the designer can make it but what is most important is if the participant can apply it to their daily task. Because our participants are firefighters, we should have researched more information on their tasks and their step by step protocols for the different scenarios they are in order to identify where in the protocol could we help firefighters become aware of what is going on around them or even also what is going on self-wise. Overall though, I think our team was able to utilize the feedback we had received both from our professors and target participants in order to fix the mistakes we made along the way. We learned that the design process is not as clean cut or mirrored as a one road path. Instead, there are different stages and sometimes you even have to go backwards and reassess ourselves through the goal ladder we learned while reading the book assigned for this class.

## 11. ACKNOWLEDGMENTS

Our thanks to Maribeth Gandy Coleman and Clint Zeagler for their instruction and help within this project and in class User Interface Design.

## 12. REFERENCES

[1] Norman, Donald A. The Design of Everyday Things. Basic Books, 2013.

[2] Tang, J. *Beginning Google Glass Development.* Apress. New York City. 2014.

[3] Van Eman II, R. L. *Finding Our Way In The Dark: A Firefighters Augmented Reality Heads-Up-Device*. 2015.

[4] Jiang, X., Hong, J. I., Takayama, L. A., & Landay, J. A. (2004, April). Ubiquitous computing for firefighters: Field studies and prototypes of large displays for incident command. In *Proceedings of the SIGCHI conference on Human factors in computing systems* (pp. 679-686).

[5] Snydal, J., Van Pelt, A., & Wilson, J. (2005). FireEye: Needs and Usability Assessment of a Head-Mounted Display for Firefighters. *University of California Berkeley*